\documentclass[twocolumn]{aastex631}

\usepackage{CJK}
\usepackage{verbatim} 

\def\lapp{\ifmmode\stackrel{<}{_{\sim}}\else$\stackrel{<}{_{\sim}}$\fi}
\def\gapp{\ifmmode\stackrel{>}{_{\sim}}\else$\stackrel{>}{_{\sim}}$\fi}

\begin{document}
\begin{CJK*}{UTF8}{gkai}

\title{Constraints on the fractional changes of the fundamental constants at a look-back time of 2.5 Myrs}

\author{Renzhi Su (苏仁智)}
\email{surenzhi@zhejianglab.com}
\affiliation{Research Center for Astronomical Computing, Zhejiang Laboratory, Hangzhou 311100, China}

\author{Tao An}
\affiliation{Shanghai Astronomical Observatory, Chinese Academy of Sciences, 80 Nandan Road, Shanghai 200030, China}
\affiliation{Xinjiang Astronomical Observatory, CAS, 150 Science 1-Street, Urumqi, Xinjiang 830011, P.R. China}
\affiliation{Guizhou Radio Astronomical Observatory, Guizhou University, 550000, Guiyang, China}

\author{Stephen, J., Curran}
\affiliation{School of Chemical and Physical Sciences, Victoria University of Wellington, PO Box 600, Wellington 6140, New Zealand}

\author[0000-0003-4961-6511]{Michael P. Busch}
\affiliation{Department of Astronomy \& Astrophysics, University of California, San Diego, 9500 Gilman Drive, La Jolla, CA 92093, USA}

\author{Minfeng Gu}
\affiliation{Shanghai Astronomical Observatory, Chinese Academy of Sciences, 80 Nandan Road, Shanghai 200030, China}

\author{Di Li}
\affiliation{Department of Astronomy, Tsinghua University, Beijing 100084, China}
\affiliation{National Astronomical Observatories, Chinese Academy of Sciences, Datun Road A20, Beijing, People’s Republic of China}
\affiliation{Research Center for Astronomical Computing, Zhejiang Laboratory, Hangzhou 311100, China}



\begin{abstract}
The quantum nature of gravity remains one of the greatest mysteries of modern physics, with many unified theories predicting variations in fundamental constants across space and time. Here we present precise measurements of these variations at galactic dynamical timescales - a critical but previously unexplored regime. Using simultaneous observations of H \textsc{i} and OH lines in M31, we probe potential variations of fundamental constants at a lookback time of 2.5 million years. We obtained $\Delta(\mu\alpha^2g_p^{0.64})/(\mu\alpha^2g_p^{0.64}) < 3.6 \times 10^{-6}$, with complementary constraints on $\Delta(\mu\alpha^2)/(\mu\alpha^2) < 4.6 \times 10^{-3}$, and $\Delta g_p/g_p < 7.2 \times 10^{-3}$, where $\alpha$ is the fine structure constant, $\mu$ is the proton-electron mass ratio, and $g_p$ is the proton $g$-factor. 
These results bridge the gap between laboratory tests and cosmological observations, providing unique insights into the coupling between local dynamics and fundamental physics. Our findings challenge theories predicting significant variations over galactic timescales, while demonstrating a powerful new probe of quantum gravity models.

\end{abstract}



\section{Introduction} \label{sec:intro}

The constancy of fundamental physical parameters stands as both a cornerstone assumption of modern physics and a crucial testing ground for quantum gravity theories. While these parameters appear fixed in everyday experiments, theoretical frameworks attempting to unify quantum mechanics and gravity --- from string theory to loop quantum gravity --- generally predict these ``constants'' may vary across space and time \citep[e.g.][]{marciano1984,damour1994,li1998,uzan2024}. The detection or constraint of such variations would provide an essential observational window into physics beyond the Standard Model, potentially revealing the quantum nature of gravity itself.  The amplitude and characteristic timescales of these variations remain key open questions. Different theoretical models predict distinct temporal behaviors: monotonic evolution with cosmic expansion \citep{damour1994}; oscillatory variations coupled to dark matter fields \citep{arvanitaki2015}; stochastic variations due to inflation \citep{peter2009,uzan2024}.

The values of low-energy coupling constants — the fine structure constant ($\alpha$), the proton-electron mass ratio ($\mu$), and the proton $g$-factor ($g_p$) — serve as sensitive probes of these potential variations. These constants affect atomic and molecular transitions differently, providing multiple observables.

Contemporary measurements span a large range of timescales. Laboratory atomic clock comparisons have achieved remarkable precision of $\sim 10^{-17}$ per year in probing variations at human timescales \citep{prestage1995,rosenband2008}. Extending to geological timescales, studies of the Oklo natural nuclear reactor  provide constraints reaching back approximately 2 Gyr into cosmic history \citep{Shlyakhter1976,damour1996}.

Astronomical observations have made particularly significant contributions across various look-back times \citep[e.g.][]{webb1999,webb2001,darling2004,srianand2004,molaro2008,curran2011,bagdonaite2013,kanekar2011,kanekar2018,muller2021,murphy2022_1,wei2024}.  For example, by measuring the stellar absorption lines towards 17 stars within 50 parsecs from Earth, $\alpha$ was not found to vary with an uncertainty of $1.2\times10^{-8}$ at a timescale of hundreds years \citep{murphy2022}. At longer timescales, The many-multiplet method applied to 143 absorption spectra from the Keck telescope's HIRES spectrograph spanning redshifts $0.2 < z < 4.2$ suggested a significant variation of $\Delta \alpha / \alpha = (0.57 \pm 0.11) \times 10^{-5}$ \citep{murphy2004}. Subsequent analysis with additional 154 absorbers from Very Large Telescope (VLT) found a possible spatial dipole model for the variation of $\alpha$ across the sky \citep{king2012}. Recently, similar measurements have been expanded to reshift up to $z$ $\sim$ 9.5, corresponding to a look-back time of 13.2 Gyr \citep[][]{wilczynska2020,jiang2024,wang2024}.

Radio astronomical techniques have provided independent constraints: comparison of NH$_3$ inversion lines with rotational transitions of HCO$^+$ and HCN yielded $|\Delta \mu / \mu| < 1.8 \times 10^{-6}$ at 95\% confidence \citep{murphy2008}. Using sensitive Green Bank Telescope (GBT) observations, \citet{kanekar2012} constrained $\Delta F / F = (-5.2 \pm 4.3) \times 10^{-6}$, where $F = g_p(\mu \alpha^2)^{1.57}$, by comparing \mbox{H\,{\sc i}} and OH 18-cm main line redshifts.

More recent studies continue to measure the variations of fundamental constants using improved methods, such as artificial intelligence \citep{bainbridge2017,lee2021}, or cutting-edge instruments, like James Webb Space Telescope \citep{wang2024,jiang2024} and ESPRESSO \citep[Echelle SPectrograph for Rocky Exoplanets and Stable Spectroscopic Observations;][]{pepe2021} installed on the VLT.

Despite these extensive efforts spanning laboratory, geological, and astronomical approaches, there remains a significant gap in our understanding at timescales of $10^6\sim10^7$ years — particularly valuable for testing theories that predict coupling between fundamental constants and local gravitational dynamics \citep[e.g.][]{damour1994,khoury2004}, as it is consistent with the timescale for star formation \citep[e.g.][]{mckee2007,clark2012}. In this paper, we report constraints on the values of fundamental constants at a look-back time of $\sim$ 2.5 Myrs, by comparing extremely sensitive radio observations of HI and 18cm OH in the nearby galaxy M31. Our measurements at 2.5 Myr lookback time provide a unique probe of this previously unexplored regime, complementing existing constraints and testing theoretical predictions about the interplay between fundamental physics and galactic dynamics.

\section{Data and Methodology}

\subsection{Observations}

Spectroscopic observations of M31, at a distance of 2.5 million light-years \citep{vilardell2010}, were conducted using the GBT, targeting two positions approximately 62\arcmin{} ($\sim$13 kpc) from the galactic nucleus \citep{busch2024}. Position 1 is located at $\alpha = 0^{h}39^{m}32^{s}$, $\delta = +40^\circ26'$ (J2000) and Position 2 at $\alpha = 0^{h}40^{m}24^{s}$, $\delta = +40^\circ28'$ (J2000). These locations were strategically chosen to probe regions where both atomic and molecular gas are expected to coexist based on previous \mbox{H\,{\sc i}} M31 surveys. The observations simultaneously covered four spectral windows encompassing the \mbox{H\,{\sc i}} 21-cm line (1420.4 MHz) and the OH 18-cm lines (1612, 1665, 1667, and 1720 MHz). To maximize on-source exposure time, in-band frequency switching by $\pm2$ MHz was employed. The total integration time was approximately 21.26 and 37.30 hours for Position 1 and Position 2.

\subsection{Data and Quality Assessment}
We directly acquire the data from \cite{busch2024}. For details of the observations and data reduction, please refer to \cite{busch2024}. The data processing included careful conversion to the Local Standard of Rest (LSR) frame, accounting for time-dependent velocity corrections. Unfortunately, the OH 1612 MHz line was severely influenced by radio frequency interference (RFI), rendering unusable spectrum. The final reduced spectra revealed significant detections ($>5\sigma$) of both \mbox{H\,{\sc i}} and OH main lines (1665, 1667 MHz) at Position 1, while Position 2 yielded only \mbox{H\,{\sc i}} detection. Given our requirement for comparative analysis between species, we focus our subsequent analysis on Position 1 from which the final spectra achieved root-mean-square noise levels of 4 mK for \mbox{H\,{\sc i}} (at 1 km~s$^{-1}$ resolution) and 0.4 mK for OH (at 3.88 km~s$^{-1}$ resolution).

\subsection{Spectral Line Analysis}

Critical to our methodology is the precise measurement of relative velocities between \mbox{H\,{\sc i}} and OH transitions. We simultaneously modeled all detected lines while accounting for their known physical relationships. The OH 1665 and 1667 MHz lines are fitted with single Gaussian profiles, constrained by their theoretical intensity ratio of 5:9 under Local Thermodynamic Equilibrium (LTE) condition. The observed consistency with this ratio provides strong support for the LTE assumption and suggests the OH emission originates from a well-mixed gas phase.

The \mbox{H\,{\sc i}} profile exhibits more complex structure, requiring three Gaussian components to fit. One component exhibits velocity width consistent with that of OH lines within uncertainties. Therefore, in our final fit, we constrained the \mbox{H\,{\sc i}} component's width to match the OH lines, reflecting their likely spatial coexistence within the same gas structures. To quantify potential fundamental constant variations, we introduce two velocity offset parameters: $V_{01}$ measuring the relative velocity shift between the OH 1665 and 1667 MHz lines, and $V_{02}$ capturing any relative velocity offset between \mbox{H\,{\sc i}} and OH 1667 MHz transitions.

\subsection{Statistical Treatment and Error Analysis}

\begin{figure*}
	\includegraphics[width=\textwidth]{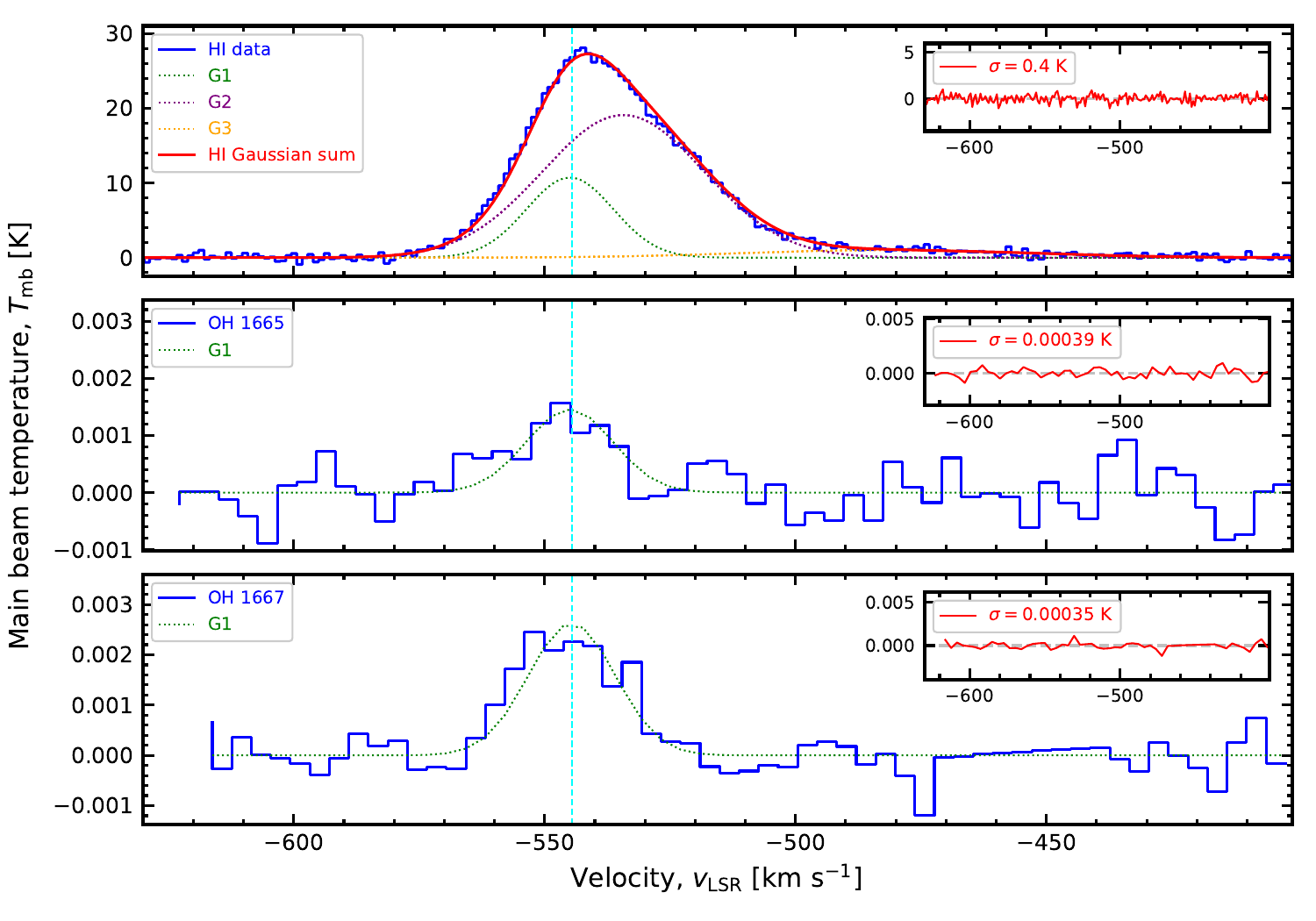}
    \caption{Gaussian fitting of the \mbox{H\,{\sc i}} and OH 18-cm lines. The blue lines are observed spectra while dotted lines represent individual Gaussian components; the red line in \mbox{H\,{\sc i}} spectrum is the sum of fitted components; the vertical dashed lines denoted the velocity at -544.53 $\rm km\,s^{-1}$; the insets show the residuals.  }
    \label{fig:spec_fit}
\end{figure*}

A particular challenge in our analysis arises from the exceptionally high signal-to-noise ratio (SNR $\approx 13800$) of the \mbox{H\,{\sc i}} data, which reveals fine spectral structure beyond the scope of our phenomenological model. We address this through a statistical approach, introducing noise of 0.4 K to the \mbox{H\,{\sc i}} spectrum. This effectively down-weights over-constrained regions while preserving the overall profile shape. We obtained the best-fit parameters of the \mbox{H\,{\sc i}} and OH lines through Monte Carlo simulations. We simultaneously fitted \mbox{H\,{\sc i}} and OH spectra for 10, 000 times. Each time, we added Gaussian random noises, characterized by the noise value of \mbox{H\,{\sc i}} and OH spectra, to the best Gaussian fits of \mbox{H\,{\sc i}} and OH lines, respectively. Finally, we took the averages as the best fitted parameters and used the standard deviations as associated uncertainties. Fits gave $\chi^2_\nu = 1.06\pm0.07$, where 1.06 is the mean $\chi^{2}$ and 0.07 is its associated standard deviation, indicating an appropriate balance between model complexity and fidelity to data.  The derived parameters are presented in Table~\ref{tab:fit_results} and a representative fit is shown in Figure~\ref{fig:spec_fit}.

Our comprehensive systematic error analysis considers multiple potential sources of uncertainty. These include velocity field variations in M31's disk, instrumental frequency registration accuracy, uncertainties in laboratory line frequencies and in the Earth's motion. Following established methodology from previous study \citep{kanekar2005}, we adopt a conservative systematic velocity uncertainty of 1.2 km~s$^{-1}$ between \mbox{H\,{\sc i}} and OH species, although the observed matched line widths suggest potentially better alignment in this specific case.

\begin{table*}
\caption{The derived parameters for the \mbox{H\,{\sc i}} and OH lines. During the fit, the velocity centers and widths of  \mbox{H\,{\sc i}}'s G1 component, OH 1665, and OH 1667 lines were tied together; the amplitude ratio between OH 1667 and OH 1665 lines was fixed to be 9:5. The fitted velocities, which account for the velocity evolution, are $V_{01}=-1.03\pm2.29\,\rm km\,s^{-1}$ and $V_{02}=-0.42\pm0.22\,\rm km\,s^{-1}$.}
\label{tab:fit_results}
\begin{tabular}{lclcc}
 \hline
Line& Name of Gaussian component  &Amplitude & Center & FWHM \\
&  &(K) & ($\rm km\,s^{-1}$)&($\rm km\,s^{-1}$)  \\
 \hline
$\mbox{H\,{\sc i}}$   & G1& 11.88$\pm$1.23  & ($-$544.53$\pm$0.24)-(0.42$\pm$0.22) & 20.70$\pm$0.83 \\        
                      & G2& 18.01$\pm$0.91 & $-$533.63$\pm$0.94 & 37.82$\pm$1.11\\
                      & G3& 1.09$\pm$0.13 & $-$487.54$\pm$8.79 & 65.48$\pm$17.55\\
\hline
OH 1665   & G1& (0.00261$\pm$0.00018)$\times$5/9 & ($-$544.53$\pm$0.24)-(1.03$\pm$2.29) & 20.70$\pm$0.83  \\
\hline
OH 1667   & G1& 0.00261$\pm$0.00018 & $-$544.53$\pm$0.24 &20.70$\pm$0.83  \\

\hline
 \end{tabular}
 \end{table*}

\section{Measuring the fractional change of the fundamental constants} \label{sec:mc}

\subsection{Theoretical Framework}

The rest-frame frequencies of atomic and molecular transitions exhibit distinct dependencies on fundamental physical constants \citep[e.g.][]{chengalur2003,curran2004}. Following the formalism developed by \citet{chengalur2003}\footnote{Note that $\mu$ is defined as electron-proton mass ratio in \cite{chengalur2003}, whereas it is the proton-electron mass ratio in this paper.}, we can express the relative frequency shifts between different transitions in terms of variations in the fine structure constant, the proton-electron mass ratio, and the proton g-factor. These dependencies arise from the underlying quantum mechanical structure of the transitions and provide a sensitive probe of potential variations in these constants.

\subsection{Analysis Methodology}

We employ two independent comparisons to constrain variations in the fundamental constants. First, comparing the measured redshifts between the \mbox{H\,{\sc i}} line and the sum frequency of OH main lines ($\nu_{1665}+\nu_{1667}$), we obtain:

\begin{equation}\label{equ:compa_oh_main_hi}
\frac{\Delta z_{1}}{1+z_{1}} = 1.57\frac{\Delta \mu}{\mu}+3.14\frac{\Delta \alpha}{\alpha}+\frac{\Delta g_p}{g_p}=\frac{\Delta (\mu^{1.57}\alpha^{3.14}g_p)}{\mu^{1.57}\alpha^{3.14}g_p}.
\end{equation}
where $\Delta z_{1} = z(\nu_{1667}+\nu_{1665})-z$(H{\sc i}) represents the redshift difference, $z_{1} = [z(\nu_{1667}+\nu_{1665})+z$(H{\sc i})]/2 is the mean redshift.

Second, we analyze the frequency difference between OH main lines relative to their sum frequency:
\begin{equation}\label{equ:compa_oh_main}
\frac{\Delta z_{2}}{1+z_{2}} = 0.13\frac{\Delta \mu}{\mu}+0.26\frac{\Delta \alpha}{\alpha}+\frac{\Delta g_p}{g_p}=\frac{\Delta (\mu^{0.13}\alpha^{0.26}g_p)}{\mu^{0.13}\alpha^{0.26}g_p},
\end{equation}
where $\Delta z_{2} = z(\nu_{1667}+\nu_{1665})-z(\nu_{1667}-\nu_{1665})$ is the difference of measured redshifts between $\nu_{1667}+\nu_{1665}$ and  $\nu_{1667}-\nu_{1665}$, $z_{2} = [z(\nu_{1667}+\nu_{1665})-z(\nu_{1667}-\nu_{1665})]/2$ is the mean redshift.

To measure the $z(\nu_{1667}+\nu_{1665})$ and $z(\nu_{1667}-\nu_{1665})$, we first converted the velocity to frequency using the equation:
\begin{equation}\label{equ:f_to_v}
V=\frac{\nu_{0}-\nu}{\nu}\times c
\end{equation}
, where $V$ is observed velocity, $\nu_{0}$ is rest-frame frequency, $\nu$ is observed frequency, and $c$ is the velocity of light. We then obtained $z(\nu_{1667}+\nu_{1665})$ = $(\nu_{0,1667}+\nu_{0,1665})/(\nu_{1667}+\nu_{1665})$ - 1 and  $z(\nu_{1667}-\nu_{1665})$ = $(\nu_{0,1667}-\nu_{0,1665})/(\nu_{1667}-\nu_{1665})$ - 1, where $\nu_{1667}$ and $\nu_{1665} $ are observed frequencies of OH 1667 and OH 1665 lines and can be determined using the Equation \ref{equ:f_to_v} and the fitted velocity, $\nu_{0,1667}$ = $1,667,358,996 \pm 4$ Hz and  $\nu_{0,1665}$ = $1,665,401,803 \pm 12$ Hz are rest-frame frequencies of OH 1667 and OH 1665 lines \citep{hudson2006}.

\subsection{Results}

By applying the Equations \ref{equ:compa_oh_main_hi} and \ref{equ:compa_oh_main} to the `G1' component identified in our spectral analysis (Table \ref{tab:fit_results}):  

\begin{equation}\label{equ:resuls}
 \left\{
  \begin{array}{ll}
   1.57\frac{\Delta \mu}{\mu}+3.14\frac{\Delta \alpha}{\alpha}+\frac{\Delta g_p}{g_p}= (0.3\pm4.0)\times10^{-6}\\
    0.13\frac{\Delta \mu}{\mu}+0.26\frac{\Delta \alpha}{\alpha}+\frac{\Delta g_p}{g_p}=(-3.0\pm6.6)\times10^{-3},
  \end{array}
  \right.
\end{equation}
from which we obtain:
\begin{equation}\label{equ:final_result1}
\frac{\Delta (\mu\alpha^{2}g_{p}^{0.64})}{\mu\alpha^{2}g_{p}^{0.64}} = (0.2\pm2.5)\times10^{-6},
\end{equation}

\begin{equation}\label{equ:final_result2}
\frac{\Delta (\mu\alpha^{2})}{\mu\alpha^{2}} = (2.1\pm4.6)\times10^{-3},
\end{equation}
and 
\begin{equation}\label{equ:final_result3}
\frac{\Delta g_{p}}{g_{p}} = (-3.3\pm7.2)\times10^{-3}.
\end{equation}

These results provide independent constraints on three different combinations of fundamental constants at a look-back time of 2.5 Myr. The tightest constraint, on the combination $\mu\alpha^{2}g_{p}^{0.64}$, reaches a precision of $\sim10^{-6}$, comparable to the best known astronomical measurement on the same combination at the other epoch \citep{kanekar2012}. When expressed as rates of change, these translate to limits of 
\[
\frac{\Delta(\mu \alpha^2 g_p^{0.64})}{(\mu \alpha^2 g_p^{0.64})} < 1.0 \times 10^{-12} \, \text{yr}^{-1},
\]
\[
\frac{\Delta(\mu \alpha^2)}{(\mu \alpha^2)} < 1.8 \times 10^{-9} \, \text{yr}^{-1},
\]
\[
\frac{\Delta g_p}{g_p} < 2.9 \times 10^{-9} \, \text{yr}^{-1},
\]
offering a new probe of potential variations at galactic dynamical timescales.

\section{Discussion} 

The robustness of constraints on fundamental constant variations critically depends on our understanding and control of systematic effects. The observations were carried out with the GBT and the data were recorded in topocentric frequency without correction for the Earth's motion on time. The conversion to the LSR frame was performed in the data reduction after averaging each cycle spectrum. In principle, all observational elements involved in the conversion, such as the model of Earth motion, the velocity difference caused by Earth motion in each observing cycle, and the rotation of the disk of M31, could influence the accuracy of our results. However, the simultaneous observation of \mbox{H\,{\sc i}} and OH lines eliminates many time-dependent systematic effects. This simultaneity ensures that any temporal variations in observing conditions affect all spectral features equally, preserving their relative velocity differences between lines which are crucial for our analysis.

The uncertainties, $\sim100$ Hz for the OH main lines \citep{ter1972}, of the laboratory rest-frame line frequencies can contribute the uncertainties of the final results. These have been considered in our calculations and have been included in our results of Equations \ref{equ:final_result1}, \ref{equ:final_result2}, and \ref{equ:final_result3}, but have negligible effects. In contrast, instrumental frequency registration accuracy, within a few Hz, is much smaller than the uncertainties of the laboratory rest-frame line frequencies and hence has no measurable impact on our results.

A primary concern in any spectral comparison study is the potential physical velocity offset between different species. Previous studies \citep{kanekar2005,kanekar2012} have shown that velocity offsets between \mbox{H\,{\sc i}} and molecular tracers like HCO$^+$ typically do not exceed 1.2 km~s$^{-1}$ in the Galaxy \citep{Drinkwater1998, Liszt2000}. we conservatively adopt this value as a systematic uncertainty, which contributes an additional uncertainty of $4.0 \times 10^{-6}$ in redshift. By applying this error, we obtain:

\[
\frac{\Delta (\mu\alpha^{2}g_{p}^{0.64})}{\mu\alpha^{2}g_{p}^{0.64}} < 3.6\times10^{-6} \Rightarrow < 1.4 \times 10^{-12} \, \text{yr}^{-1},
\]
\[
\frac{\Delta (\mu\alpha^{2})}{\mu\alpha^{2}}< 4.6\times10^{-3} \Rightarrow < 1.8 \times 10^{-9} \, \text{yr}^{-1},
\]
and
\[
\frac{\Delta g_{p}}{g_{p}} < 7.2\times10^{-3} \Rightarrow < 2.9 \times 10^{-9} \, \text{yr}^{-1}.
\]

Our results at a look-back time of 2.5 Myr provide important complementary constraints to existing measurements from laboratory measurements ($\sim$years) and high-redshift astronomical observations ($\sim$Gyr) \citep[e.g.][]{murphy2004,king2012,milakovic2021,murphy2022_1}. The precision achieved for the combination $\mu \alpha^2 g_p^{0.64}$ ($\sim 10^{-6}$) demonstrates the power of simultaneous HI and OH observations in testing the constancy of fundamental constants. While this single measurement cannot fully constrain specific theoretical models, it adds a valuable data point at an intermediate timescale. Future observations of multiple positions and galaxies could help build a more complete picture of potential variations across different physical environments and timescales.

\section{Summary and Conclusions}

Using simultaneous observations of \mbox{H\,{\sc i}} 21-cm and OH 18-cm lines towards M31, we have conducted a precise test for variations in fundamental physical constants at a look-back time of 2.5 Myr. Our most stringent constraint reaches a precision of $3.6 \times 10^{-6}$ for the combination of constants $\mu \alpha^2 g_p^{0.64}$, with complementary bounds on $\mu \alpha^2$ and $g_p$ at the $10^{-3}$ level. When expressed as rates of change, these translate to upper limits of order $10^{-12}$ to $10^{-9}$ yr$^{-1}$.

These measurements are significant as they probe a previously unexplored timescale, bridging the gap between laboratory experiments and cosmological observations. While this timescale does not directly probe full galactic orbital periods, it provides important constraints on shorter-term variations that might arise from local physical processes or cosmological evolution. The simultaneous measurement of multiple transitions allows us to eliminate some systematic effects, which increases the robustness of the result. The precision achieved in our measurements demonstrates the power of radio astronomical techniques for testing fundamental physics. Future observations with next-generation facilities like the Square Kilometre Array \citep[SKA;][]{dewdney2009} could greatly improve these constraints.

\begin{acknowledgments}

This work is partly supported by National Natural Science Foundation of China (Grant No. 11988101). RSU acknowledges the support from the China Postdoctoral Science Foundation (Grant No. 2024M752979). This work is partly supported by National SKA Program of China (2022SKA0120102).  M.P.B. is supported by an NSF Astronomy and Postdoctoral Fellowship under the award AST-2202373. TA acknowledge the support from the Xinjiang Tianchi Talent Program and the FAST special Program (NSFC 12041301).  MFG is supported by the National Science Foundation of China (grant 12473019), the China Manned Space Project with No. CMSCSST-2021-A06, the National SKA Program of China (Grant No. 2022SKA0120102), and the Shanghai Pilot Program for Basic Research-Chinese Academy of Science, Shanghai Branch (JCYJ-SHFY-2021-013).
\end{acknowledgments}

%





\bibliography{vari_constants}{}

\begin{thebibliography}{}
\expandafter\ifx\csname natexlab\endcsname\relax\def\natexlab#1{#1}\fi
\providecommand{\url}[1]{\href{#1}{#1}}
\providecommand{\dodoi}[1]{doi:~\href{http://doi.org/#1}{\nolinkurl{#1}}}
\providecommand{\doeprint}[1]{\href{http://ascl.net/#1}{\nolinkurl{http://ascl.net/#1}}}
\providecommand{\doarXiv}[1]{\href{https://arxiv.org/abs/#1}{\nolinkurl{https://arxiv.org/abs/#1}}}

\bibitem[{{Arvanitaki} {et~al.}(2015){Arvanitaki}, {Huang}, \& {Van Tilburg}}]{arvanitaki2015}
{Arvanitaki}, A., {Huang}, J., \& {Van Tilburg}, K. 2015, \prd, 91, 015015, \dodoi{10.1103/PhysRevD.91.015015}

\bibitem[{{Bagdonaite} {et~al.}(2013){Bagdonaite}, {Jansen}, {Henkel}, {Bethlem}, {Menten}, \& {Ubachs}}]{bagdonaite2013}
{Bagdonaite}, J., {Jansen}, P., {Henkel}, C., {et~al.} 2013, Science, 339, 46, \dodoi{10.1126/science.1224898}

\bibitem[{{Bainbridge} \& {Webb}(2017)}]{bainbridge2017}
{Bainbridge}, M.~B., \& {Webb}, J.~K. 2017, \mnras, 468, 1639, \dodoi{10.1093/mnras/stx179}

\bibitem[{{Busch}(2024)}]{busch2024}
{Busch}, M.~P. 2024, \apj, 967, 148, \dodoi{10.3847/1538-4357/ad3af6}

\bibitem[{{Chengalur} \& {Kanekar}(2003)}]{chengalur2003}
{Chengalur}, J.~N., \& {Kanekar}, N. 2003, \prl, 91, 241302, \dodoi{10.1103/PhysRevLett.91.241302}

\bibitem[{{Clark} {et~al.}(2012){Clark}, {Glover}, {Klessen}, \& {Bonnell}}]{clark2012}
{Clark}, P.~C., {Glover}, S. C.~O., {Klessen}, R.~S., \& {Bonnell}, I.~A. 2012, \mnras, 424, 2599, \dodoi{10.1111/j.1365-2966.2012.21259.x}

\bibitem[{{Curran} {et~al.}(2004){Curran}, {Kanekar}, \& {Darling}}]{curran2004}
{Curran}, S.~J., {Kanekar}, N., \& {Darling}, J.~K. 2004, \nar, 48, 1095, \dodoi{10.1016/j.newar.2004.09.004}

\bibitem[{{Curran} {et~al.}(2011){Curran}, {Tanna}, {Koch}, {Berengut}, {Webb}, {Stark}, \& {Flambaum}}]{curran2011}
{Curran}, S.~J., {Tanna}, A., {Koch}, F.~E., {et~al.} 2011, \aap, 533, A55, \dodoi{10.1051/0004-6361/201117457}

\bibitem[{{Damour} \& {Dyson}(1996)}]{damour1996}
{Damour}, T., \& {Dyson}, F. 1996, Nuclear Physics B, 480, 37, \dodoi{10.1016/S0550-3213(96)00467-1}

\bibitem[{{Damour} \& {Polyakov}(1994)}]{damour1994}
{Damour}, T., \& {Polyakov}, A.~M. 1994, Nuclear Physics B, 423, 532, \dodoi{10.1016/0550-3213(94)90143-0}

\bibitem[{{Darling}(2004)}]{darling2004}
{Darling}, J. 2004, \apj, 612, 58, \dodoi{10.1086/422450}

\bibitem[{{Dewdney} {et~al.}(2009){Dewdney}, {Hall}, {Schilizzi}, \& {Lazio}}]{dewdney2009}
{Dewdney}, P.~E., {Hall}, P.~J., {Schilizzi}, R.~T., \& {Lazio}, T.~J.~L.~W. 2009, IEEE Proceedings, 97, 1482, \dodoi{10.1109/JPROC.2009.2021005}

\bibitem[{{Drinkwater} {et~al.}(1998){Drinkwater}, {Webb}, {Barrow}, \& {Flambaum}}]{Drinkwater1998}
{Drinkwater}, M.~J., {Webb}, J.~K., {Barrow}, J.~D., \& {Flambaum}, V.~V. 1998, \mnras, 295, 457, \dodoi{10.1046/j.1365-8711.1998.2952457.x}

\bibitem[{{Hudson} {et~al.}(2006){Hudson}, {Lewandowski}, {Sawyer}, \& {Ye}}]{hudson2006}
{Hudson}, E.~R., {Lewandowski}, H.~J., {Sawyer}, B.~C., \& {Ye}, J. 2006, \prl, 96, 143004, \dodoi{10.1103/PhysRevLett.96.143004}

\bibitem[{{Jiang} {et~al.}(2024){Jiang}, {Pan}, {Aguilar}, {Ahlen}, {Blum}, {Brooks}, {Claybaugh}, {de la Macorra}, {Dey}, {Doel}, {Fanning}, {Ferraro}, {Forero-Romero}, {Gazta{\~n}aga}, {Gontcho A Gontcho}, {Gutierrez}, {Honscheid}, {Juneau}, {Landriau}, {Le Guillou}, {Levi}, {Manera}, {Miquel}, {Moustakas}, {Mueller}, {Mu{\~n}oz-Guti{\'e}rrez}, {Myers}, {Nie}, {Niz}, {Poppett}, {Prada}, {Rezaie}, {Rossi}, {Sanchez}, {Schlafly}, {Schubnell}, {Seo}, {Sprayberry}, {Tarl{\'e}}, {Weaver}, {Zou}, \& {The DESI Collaboration}}]{jiang2024}
{Jiang}, L., {Pan}, Z., {Aguilar}, J.~N., {et~al.} 2024, \apj, 968, 120, \dodoi{10.3847/1538-4357/ad47b4}

\bibitem[{{Kanekar}(2011)}]{kanekar2011}
{Kanekar}, N. 2011, \apjl, 728, L12, \dodoi{10.1088/2041-8205/728/1/L12}

\bibitem[{{Kanekar} {et~al.}(2018){Kanekar}, {Ghosh}, \& {Chengalur}}]{kanekar2018}
{Kanekar}, N., {Ghosh}, T., \& {Chengalur}, J.~N. 2018, \prl, 120, 061302, \dodoi{10.1103/PhysRevLett.120.061302}

\bibitem[{{Kanekar} {et~al.}(2012){Kanekar}, {Langston}, {Stocke}, {Carilli}, \& {Menten}}]{kanekar2012}
{Kanekar}, N., {Langston}, G.~I., {Stocke}, J.~T., {Carilli}, C.~L., \& {Menten}, K.~M. 2012, \apjl, 746, L16, \dodoi{10.1088/2041-8205/746/2/L16}

\bibitem[{{Kanekar} {et~al.}(2005){Kanekar}, {Carilli}, {Langston}, {Rocha}, {Combes}, {Subrahmanyan}, {Stocke}, {Menten}, {Briggs}, \& {Wiklind}}]{kanekar2005}
{Kanekar}, N., {Carilli}, C.~L., {Langston}, G.~I., {et~al.} 2005, \prl, 95, 261301, \dodoi{10.1103/PhysRevLett.95.261301}

\bibitem[{{Khoury} \& {Weltman}(2004)}]{khoury2004}
{Khoury}, J., \& {Weltman}, A. 2004, \prd, 69, 044026, \dodoi{10.1103/PhysRevD.69.044026}

\bibitem[{{King} {et~al.}(2012){King}, {Webb}, {Murphy}, {Flambaum}, {Carswell}, {Bainbridge}, {Wilczynska}, \& {Koch}}]{king2012}
{King}, J.~A., {Webb}, J.~K., {Murphy}, M.~T., {et~al.} 2012, \mnras, 422, 3370, \dodoi{10.1111/j.1365-2966.2012.20852.x}

\bibitem[{{Lee} {et~al.}(2021){Lee}, {Webb}, {Carswell}, \& {Milakovi{\'c}}}]{lee2021}
{Lee}, C.-C., {Webb}, J.~K., {Carswell}, R.~F., \& {Milakovi{\'c}}, D. 2021, \mnras, 504, 1787, \dodoi{10.1093/mnras/stab977}

\bibitem[{{Li} \& {Gott}(1998)}]{li1998}
{Li}, L.-X., \& {Gott}, J.~Richard, I. 1998, \prd, 58, 103513, \dodoi{10.1103/PhysRevD.58.103513}

\bibitem[{{Liszt} \& {Lucas}(2000)}]{Liszt2000}
{Liszt}, H., \& {Lucas}, R. 2000, \aap, 355, 333

\bibitem[{{Marciano}(1984)}]{marciano1984}
{Marciano}, W.~J. 1984, \prl, 52, 489, \dodoi{10.1103/PhysRevLett.52.489}

\bibitem[{{McKee} \& {Ostriker}(2007)}]{mckee2007}
{McKee}, C.~F., \& {Ostriker}, E.~C. 2007, \araa, 45, 565, \dodoi{10.1146/annurev.astro.45.051806.110602}

\bibitem[{{Milakovi{\'c}} {et~al.}(2021){Milakovi{\'c}}, {Lee}, {Carswell}, {Webb}, {Molaro}, \& {Pasquini}}]{milakovic2021}
{Milakovi{\'c}}, D., {Lee}, C.-C., {Carswell}, R.~F., {et~al.} 2021, \mnras, 500, 1, \dodoi{10.1093/mnras/staa3217}

\bibitem[{{Molaro} {et~al.}(2008){Molaro}, {Levshakov}, {Monai}, {Centuri{\'o}n}, {Bonifacio}, {D'Odorico}, \& {Monaco}}]{molaro2008}
{Molaro}, P., {Levshakov}, S.~A., {Monai}, S., {et~al.} 2008, \aap, 481, 559, \dodoi{10.1051/0004-6361:20078864}

\bibitem[{{Muller} {et~al.}(2021){Muller}, {Ubachs}, {Menten}, {Henkel}, \& {Kanekar}}]{muller2021}
{Muller}, S., {Ubachs}, W., {Menten}, K.~M., {Henkel}, C., \& {Kanekar}, N. 2021, \aap, 652, A5, \dodoi{10.1051/0004-6361/202140531}

\bibitem[{{Murphy} {et~al.}(2022{\natexlab{a}}){Murphy}, {Berke}, {Liu}, {Flynn}, {Lehmann}, {Dzuba}, \& {Flambaum}}]{murphy2022}
{Murphy}, M.~T., {Berke}, D.~A., {Liu}, F., {et~al.} 2022{\natexlab{a}}, Science, 378, 634, \dodoi{10.1126/science.abi9232}

\bibitem[{{Murphy} {et~al.}(2008){Murphy}, {Flambaum}, {Muller}, \& {Henkel}}]{murphy2008}
{Murphy}, M.~T., {Flambaum}, V.~V., {Muller}, S., \& {Henkel}, C. 2008, Science, 320, 1611, \dodoi{10.1126/science.1156352}

\bibitem[{{Murphy} {et~al.}(2004){Murphy}, {Flambaum}, {Webb}, {Dzuba}, {Prochaska}, \& {Wolfe}}]{murphy2004}
{Murphy}, M.~T., {Flambaum}, V.~V., {Webb}, J.~K., {et~al.} 2004, in Astrophysics, Clocks and Fundamental Constants, ed. S.~G. {Karshenboim} \& E.~{Peik}, Vol. 648, 131--150, \dodoi{10.1007/978-3-540-40991-5_9}

\bibitem[{{Murphy} {et~al.}(2022{\natexlab{b}}){Murphy}, {Molaro}, {Leite}, {Cupani}, {Cristiani}, {D'Odorico}, {G{\'e}nova Santos}, {Martins}, {Milakovi{\'c}}, {Nunes}, {Schmidt}, {Pepe}, {Rebolo}, {Santos}, {Sousa}, {Zapatero Osorio}, {Amate}, {Adibekyan}, {Alibert}, {Allende Prieto}, {Baldini}, {Benz}, {Bouchy}, {Cabral}, {Dekker}, {Di Marcantonio}, {Ehrenreich}, {Figueira}, {Gonz{\'a}lez Hern{\'a}ndez}, {Landoni}, {Lovis}, {Lo Curto}, {Manescau}, {M{\'e}gevand}, {Mehner}, {Micela}, {Pasquini}, {Poretti}, {Riva}, {Sozzetti}, {Mascare{\~n}o}, {Udry}, \& {Zerbi}}]{murphy2022_1}
{Murphy}, M.~T., {Molaro}, P., {Leite}, A. C.~O., {et~al.} 2022{\natexlab{b}}, \aap, 658, A123, \dodoi{10.1051/0004-6361/202142257}

\bibitem[{{Pepe} {et~al.}(2021){Pepe}, {Cristiani}, {Rebolo}, {Santos}, {Dekker}, {Cabral}, {Di Marcantonio}, {Figueira}, {Lo Curto}, {Lovis}, {Mayor}, {M{\'e}gevand}, {Molaro}, {Riva}, {Zapatero Osorio}, {Amate}, {Manescau}, {Pasquini}, {Zerbi}, {Adibekyan}, {Abreu}, {Affolter}, {Alibert}, {Aliverti}, {Allart}, {Allende Prieto}, {{\'A}lvarez}, {Alves}, {Avila}, {Baldini}, {Bandy}, {Barros}, {Benz}, {Bianco}, {Borsa}, {Bourrier}, {Bouchy}, {Broeg}, {Calderone}, {Cirami}, {Coelho}, {Conconi}, {Coretti}, {Cumani}, {Cupani}, {D'Odorico}, {Damasso}, {Deiries}, {Delabre}, {Demangeon}, {Dumusque}, {Ehrenreich}, {Faria}, {Fragoso}, {Genolet}, {Genoni}, {G{\'e}nova Santos}, {Gonz{\'a}lez Hern{\'a}ndez}, {Hughes}, {Iwert}, {Kerber}, {Knudstrup}, {Landoni}, {Lavie}, {Lillo-Box}, {Lizon}, {Maire}, {Martins}, {Mehner}, {Micela}, {Modigliani}, {Monteiro}, {Monteiro}, {Moschetti}, {Murphy}, {Nunes}, {Oggioni}, {Oliveira}, {Oshagh}, {Pall{\'e}}, {Pariani}, {Poretti}, {Rasilla}, {Rebord{\~a}o}, {Redaelli}, {Santana Tschudi},
  {Santin}, {Santos}, {S{\'e}gransan}, {Schmidt}, {Segovia}, {Sosnowska}, {Sozzetti}, {Sousa}, {Span{\`o}}, {Su{\'a}rez Mascare{\~n}o}, {Tabernero}, {Tenegi}, {Udry}, \& {Zanutta}}]{pepe2021}
{Pepe}, F., {Cristiani}, S., {Rebolo}, R., {et~al.} 2021, \aap, 645, A96, \dodoi{10.1051/0004-6361/202038306}

\bibitem[{Peter \& Uzan(2009)}]{peter2009}
Peter, P., \& Uzan, J.-P. 2009, Primordial Cosmology (Oxford; New York: Oxford University Press)

\bibitem[{{Prestage} {et~al.}(1995){Prestage}, {Tjoelker}, \& {Maleki}}]{prestage1995}
{Prestage}, J.~D., {Tjoelker}, R.~L., \& {Maleki}, L. 1995, \prl, 74, 3511, \dodoi{10.1103/PhysRevLett.74.3511}

\bibitem[{{Rosenband} {et~al.}(2008){Rosenband}, {Hume}, {Schmidt}, {Chou}, {Brusch}, {Lorini}, {Oskay}, {Drullinger}, {Fortier}, {Stalnaker}, {Diddams}, {Swann}, {Newbury}, {Itano}, {Wineland}, \& {Bergquist}}]{rosenband2008}
{Rosenband}, T., {Hume}, D.~B., {Schmidt}, P.~O., {et~al.} 2008, Science, 319, 1808, \dodoi{10.1126/science.1154622}

\bibitem[{{Shlyakhter}(1976)}]{Shlyakhter1976}
{Shlyakhter}, A.~I. 1976, \nat, 264, 340, \dodoi{10.1038/264340a0}

\bibitem[{{Srianand} {et~al.}(2004){Srianand}, {Chand}, {Petitjean}, \& {Aracil}}]{srianand2004}
{Srianand}, R., {Chand}, H., {Petitjean}, P., \& {Aracil}, B. 2004, \prl, 92, 121302, \dodoi{10.1103/PhysRevLett.92.121302}

\bibitem[{{Ter Meulen} \& {Dymanus}(1972)}]{ter1972}
{Ter Meulen}, J.~J., \& {Dymanus}, A. 1972, \apjl, 172, L21, \dodoi{10.1086/180882}

\bibitem[{{Uzan}(2024)}]{uzan2024}
{Uzan}, J.-P. 2024, arXiv e-prints, arXiv:2410.07281, \dodoi{10.48550/arXiv.2410.07281}

\bibitem[{{Vilardell} {et~al.}(2010){Vilardell}, {Ribas}, {Jordi}, {Fitzpatrick}, \& {Guinan}}]{vilardell2010}
{Vilardell}, F., {Ribas}, I., {Jordi}, C., {Fitzpatrick}, E.~L., \& {Guinan}, E.~F. 2010, \aap, 509, A70, \dodoi{10.1051/0004-6361/200913299}

\bibitem[{{Wang} {et~al.}(2024){Wang}, {Lei}, {Feng}, \& {Fan}}]{wang2024}
{Wang}, Z.-F., {Lei}, L., {Feng}, L., \& {Fan}, Y.-Z. 2024, Research in Astronomy and Astrophysics, 24, 125012, \dodoi{10.1088/1674-4527/ad9198}

\bibitem[{{Webb} {et~al.}(1999){Webb}, {Flambaum}, {Churchill}, {Drinkwater}, \& {Barrow}}]{webb1999}
{Webb}, J.~K., {Flambaum}, V.~V., {Churchill}, C.~W., {Drinkwater}, M.~J., \& {Barrow}, J.~D. 1999, \prl, 82, 884, \dodoi{10.1103/PhysRevLett.82.884}

\bibitem[{{Webb} {et~al.}(2001){Webb}, {Murphy}, {Flambaum}, {Dzuba}, {Barrow}, {Churchill}, {Prochaska}, \& {Wolfe}}]{webb2001}
{Webb}, J.~K., {Murphy}, M.~T., {Flambaum}, V.~V., {et~al.} 2001, \prl, 87, 091301, \dodoi{10.1103/PhysRevLett.87.091301}

\bibitem[{{Wei} {et~al.}(2024){Wei}, {Chen}, {Wei}, {Lopez-Corredoira}, \& {Wu}}]{wei2024}
{Wei}, J.-N., {Chen}, R.-J., {Wei}, J.-J., {Lopez-Corredoira}, M., \& {Wu}, X.-F. 2024, arXiv e-prints, arXiv:2409.01554, \dodoi{10.48550/arXiv.2409.01554}

\bibitem[{{Wilczynska} {et~al.}(2020){Wilczynska}, {Webb}, {Bainbridge}, {Barrow}, {Bosman}, {Carswell}, {D{\k{a}}browski}, {Dumont}, {Lee}, {Leite}, {Leszczy{\'n}ska}, {Liske}, {Marosek}, {Martins}, {Milakovi{\'c}}, {Molaro}, \& {Pasquini}}]{wilczynska2020}
{Wilczynska}, M.~R., {Webb}, J.~K., {Bainbridge}, M., {et~al.} 2020, Science Advances, 6, eaay9672, \dodoi{10.1126/sciadv.aay9672}

\end{thebibliography}
\bibliographystyle{aasjournal}


\end{CJK*}
\end{document}